\documentstyle[prl,epsfig,graphics,aps,multicol]{revtex}

\draft
\title{Effect of initial correlations on short--time decoherence}
\author{Eric Lutz}
\address{ Sloane Physics Laboratory, Yale University, P.O. Box 208120, New Haven, CT 06520-8120, USA}
\date{\today}

\newcommand{\la}{\langle}
\newcommand{\ra}{\rangle}
\newcommand{\SC}{\scriptscriptstyle}
\newcommand{\be}{\begin{equation}}
\newcommand{\ee}{\end{equation}}
\newcommand{\bea}{\begin{eqnarray}}
\newcommand{\eea}{\end{eqnarray}}

\begin{document}

\maketitle
\begin{multicols}{2}[
\begin{abstract}
We study the effect of initial correlations on the short--time
decoherence of a particle linearly coupled to a bath of harmonic
oscillators. We analytically evaluate the attenuation coefficient
of a Schr\"odinger cat state both for a free  and a harmonically
bound particle, with and without initial thermal correlations
between the particle and the bath. While  short--time
decoherence appears to be independent of the system in the absence
of initial correlations, we find on the contrary  that, for
initial thermal correlations,  decoherence becomes system
dependent even for times much shorter than the characteristic time of the 
system. The temperature behavior of this system dependence is
discussed.
\end{abstract}
]

Environment induced decoherence plays a fundamental role in many
areas, ranging from quantum cosmology \cite{bar99} and the theory
of quantum measurement \cite{whe83} to quantum information and
quantum computing \cite{gal02}. Experimental investigations of the
decoherence process have recently been reported in
Refs.~\cite{bru96,mya00,mei00}. Environmental  decoherence can be
defined as "... the (irreversible) loss of quantum coherence of a
quantum system due to its coupling to an environment"
\cite{moh01}; it manifests itself in the dynamical suppression of
interference phenomena \cite{zur81,giu96}.  A simple way to
quantify the destruction of coherence is thus to consider the
superposition of two localized wave packets (Schr\"odinger cat
state) and to look at the decay of the interference term, as
measured for instance by the attenuation coefficient [see
Eq.~(\ref{eq7}) below]. An important characteristic of decoherence
is that it occurs on a very fast time scale, usually much shorter
than the energy dissipation scale. We note that the short--time
limit of decoherence has recently attracted a renewed interest in
the literature \cite{for01a,for01b,haa01,haa02,pri}. Interestingly, Braun, Haake and
Strunz \cite{haa01,haa02} have identified a new regime of fast decoherence
beyond the usual Golden Rule regime. In this limit of large
separations between the wave packets (interaction-dominated
decoherence), quantum decoherence appears independent of both the
system and the heat bath.

Most studies of environment induced decoherence make use of the
simplified assumption that the system and the bath are initially
uncorrelated \cite{paz99}. Then, the initial composite density
operator for system plus bath can be factorized into a product of
a system operator and a bath operator. However, in the  general,
and more realistic case, initial correlations can be present and
the latter might affect the decoherence process, especially at
very short times as first pointed out by Romero and Paz \cite{rom97}. In
fact, as recently shown by Ford, Lewis and O'Connell \cite{for01a,for01b},
initial thermal conditions  can dramatically modify the
decoherence rate. They find that the decoherence time for an
unbound particle becomes independent of the strength of the
dissipation when it is initially at the same temperature as the
bath. Our aim in this paper is to complement the discussion of the
effect of initial thermal correlations on short--time decoherence
by analytically comparing the attenuation factor of a free
particle with that of a harmonic oscillator, with and without
thermal initial correlations. As our main tool, we shall employ
the by now standard model of a system linearly coupled to a bath
of harmonic oscillators. This model is exactly solvable and the
exact reduced density operator of the quantum system is
conveniently obtained within a path integral approach
\cite{fey63}. This model has been extensively used in decoherence
studies  for the case of factorizable initial conditions, building
up on the work of Caldeira and Leggett \cite{cal83}. It has been
extended later on to correlated thermal initial conditions by
Hakim and Ambegaokar for a free particle \cite{hak85} and by
Morais Smith and Caldeira for a harmonically bound particle
\cite{mor87,mor90} (see also the work of Grabert and collaborators
\cite{gra88}). In the following we shall use the exact results
derived by Morais Smith and Caldeira \cite{mor90} to compute the
time evolution of the superposition of two Gaussian wave packets
separated by a distance $2a$ and of width $b$, $\psi(x) = c_1
\exp[-(x-a)^2/4b^2]+  c_2 \exp[-(x+a)^2/4b^2]$. The corresponding
initial  density operator, $\rho_S(x,y,0)=\psi(x)\psi^*(y)$, is
given by
 \bea \label{eq0}
 \rho_s(x,y,0) &=&|c_1|^2
 \exp\Big[-\frac{(x-a)^2+(y-a)^2}{4b^2}\Big] \nonumber\\
 &+& |c_2|^2 \exp\Big[-\frac{(x+a)^2+(y+a)^2}{4b^2}\Big]\nonumber \\
 &+&c_1c_2^* \exp\Big[-\frac{(x-a)^2+(y+a)^2}{4b^2}\Big] \nonumber \\
 &+&c_1^*c_2 \exp\Big[-\frac{(x+a)^2+(y-a)^2}{4b^2}\Big]\ .
 \eea
 We calculate the short--time limit of the attenuation coefficient for a
harmonically bound particle with and without initial thermal
correlations and compare the obtained results with those of a free
particle. We find that for initial decorrelation, the attenuation
coefficient is identical for both the free particle and the linear
oscillator, that is to say, decoherence is independent of the nature of
the system. On the other hand, for initial thermal correlations,
we find that the attenuation coefficient acquires a system
dependent part, even for times much smaller than the
characteristic time of the system. This shows that in the presence
of initial thermal correlations decoherence is system dependent.
At very high temperature, this system dependence turns out  to be
negligibly small. However, in the opposite limit of low
temperature, it becomes increasingly important.

 {\bf The model}. We consider a particle of unit mass,
moving in an external potential $V(x)$ and linearly coupled
through its position to a set of independent harmonic oscillators
with mass $m_i$ and frequency $\omega_i$. The composite
Hamiltonian is written in the form
 \bea \label{eq1}
 H&=& \frac{p^2}{2} + V(x) +
 \sum_i x \, C_i x_i + \sum_i \left(
 \frac{p^2_i}{2m_i}+\frac{1}{2}m_i\omega_i^2 x_i^2\right) \nonumber \\
 &&+ \sum_i x^2\frac{C_i^2}{2m_i \omega_i^2} \ ,
 \eea
 where the $C_i$'s are coupling constants. In the limit of infinitely many oscillators, the bath is entirely
characterized by the spectral density function, \be \label{eq2}
I(\omega) =
\frac{\pi}{2}\sum_i\frac{C_i^2}{m_i\omega_i}\,\delta(\omega-\omega_i)=
2\gamma \omega\,\theta(w-\omega_c) \ , \ee where the last equality
defines Ohmic damping with damping coefficient $\gamma$. Here
$\omega_c$ is a cutoff frequency that will  be replaced by
infinity in all convergent integrals. The reduced density operator
$\rho_S(t)$ of the system at time $t$ is obtained after tracing
out the bath degrees of freedom. In coordinate representation it
reads
 \bea
 \label{eq3}
 \rho_S(x,y,t)=&& \int dx' dy'd{\bf R} d{\bf R'}d{\bf Q'}\,
 K(x,{\bf R},t;x',{\bf R'},0)\nonumber \\
 &&\times K^*(y,{\bf R},t;y',{\bf Q'},0) \la x,{\bf R'}|\rho(0)|y',{\bf Q'}\ra\ .
 \eea
 Here ${\bf R}$, ${\bf R'}$ and ${\bf Q'}$ collectively denote the
coordinates of the bath, $K$ is the propagator and $\rho(0)$ the
initial density operator of the composite system. If we assume
that at $t=0$ the system and the bath are uncoupled and that the
latter is in thermal equilibrium, then the initial density
operator can be written as $\rho_0(x',{\bf R'};y',{\bf Q'})=
\rho_S(x',y',0)\rho_{eq}({\bf R'},{\bf Q'})$, where
$\rho_S(x',y',0)$ describes the initial state of the system and
$\rho_{eq}({\bf R'},{\bf Q'})$ is the equilibrium density operator
of the bath. On the other hand, if the system and the bath are
initially coupled and in thermal equilibrium with each other,
$\rho(0)$ cannot be factorized into a product of a system and a
bath operator anymore. Instead, we have $\rho_0(x',{\bf
R'};y',{\bf Q'}) =\rho_S(x',y',0)\rho_{eq}(x',y',{\bf R'},{\bf
Q'})$, where $\rho_{eq}(x',y',{\bf R'},{\bf Q'})$ is  now the
equilibrium density operator of the composite system. We shall
refer to these two cases as (i) uncorrelated initial conditions
and (ii) thermal initial conditions, respectively. We mention that
other correlated initial conditions have also been considered (see
Refs.~\cite{mor87,mor90,gra88}). The integrals over ${\bf R}$, ${\bf
R'}$ and ${\bf Q'}$ in Eq.~(\ref{eq3}) can now be performed
exactly, yielding
 \be
 \label{eq4}
 \rho_S(q,\xi,t) = \int dq'd\xi' \,
 J(q,\xi,t;q',\xi',0) \,\rho_S(q',\xi',0)\ ,
 \ee
 where we have
introduced the center of mass and relative coordinates $q=
(x+y)/2$ and $\xi=x-y$. The dynamics of the particle is completely
determined by the propagating function $J(q,\xi,t;q',\xi',0)$.
Equation (\ref{eq4}) can be considered as the full solution of the
master equation describing the time evolution of the dissipative
particle with (and without) thermal initial conditions. In the
diagonal case $\xi=0$, the propagator $J$ is given by (we put
$\hbar\!=\!k_B\!=\!1$ throughout the paper),
 \bea
 \label{eq5}
  J(q,0,t;q',\xi',0)= &&\frac{N(t)}{2\pi}\,
 \exp\Big[i\Big(\alpha(t)\,q'\xi'-N(t)\,q \xi'\Big)\Big]\nonumber\\
 &&\times \exp\Big[-\Big(\varepsilon \,q'^2+\Delta(t)\,
 \xi'^2\Big)\Big]\ ,
 \eea
 with
 \bea \alpha(t) &=& K(t)+ \gamma - 2\varepsilon E(t) \ ,\\
 \Delta(t) &=& C(t)-\varepsilon E(t)^2 \ ,\\
\mbox{and }\hspace{10mm}  \varepsilon &=& \frac{\beta}{2\kappa}\ .
 \eea
 The exact expressions
for the coefficients $K(t)$, $N(t)$, $C(t)$ and $E(t)$  have been
derived in Ref.~\cite{mor90} for a linear oscillator with
frequency $\omega_0$ initially in thermal equilibrium with a heat
bath at inverse temperature $\beta=1/T$. They are reproduced in
the appendix for completeness. Remarkably, the form of the
propagating function (\ref{eq5}) remains the same  for
uncorrelated initial conditions, as well as for a free particle.
The case of uncorrelated initial conditions is recovered by
setting $\varepsilon$ to zero and keeping only the first double
integral in $C_\omega(t)$, Eq.~(25), [$K(t)$ and $N(t)$ being unchanged],
while the unbound particle is obtained by letting the frequency
$\omega_0$ go to zero \cite{mor90}. The quantity $\kappa$ that appears in
Eq.~(9) is equal to the variance of the position $\la x^2
\ra_{eq}$ in equilibrium [see Eq.~(\ref{eqa2}) in  appendix A].
The factor $\varepsilon$ ($=1/2\lambda$ in the notation of
\cite{mor90}), which stems from the initial thermal correlations,
will turn out to be important in the following discussion. Its
asymptotic behavior at high ($ T\!\gg \! \omega_0$) and low
temperature ($T\! \ll \! \omega_0$) is respectively given by
$\varepsilon\sim (\beta \omega_0)^2/2 \ll 1$ and $\varepsilon\sim
\beta \omega_0 \gg 1$. The diagonal density operator of the system
at time $t$ can then be easily obtained by combining
Eqs.~(\ref{eq0}), (\ref{eq4}) and (\ref{eq5}). We find
 \bea
 \label{eq6}
 \rho_S(q,0,t) &\sim& \exp\Big[-\frac{2 b^2(N q-a \alpha)^2+
 a^2(1+8b^2 \Delta)\varepsilon}{\sigma^2}\Big]\nonumber \\
 &+&\exp\Big[-\frac{2 b^2(N q+a \alpha)^2+ a^2(1+8b^2
 \Delta)\varepsilon}{\sigma^2}\Big]
 \nonumber\\
 &+& 2 \exp\Big[- \frac{a^2(4 \Delta+2b^2\alpha^2)+2b^2N^2
 q^2}{\sigma^2}\Big]\nonumber \\
 &&\times\cos\frac{2a N q(1+2b^2 \varepsilon)}{\sigma^2} \ ,
 \eea
 where we have defined  $\sigma^2= (1+8b^2 \Delta)(1+2b^2\varepsilon)+ 4 b^4\alpha^2$
(for simplicity, we have put $c_1$ equal to $c_2$).
Equation(\ref{eq6}) is written as a sum of three terms. The first
two terms correspond to  two separately propagating wave packets,
while the third one, containing the cosine, is an interference
term. The attenuation coefficient $a(t)$ is defined as the ratio
of the factor multiplying the cosine to twice the geometric mean
of the first two terms. It follows from Eq.~(\ref{eq6}) that
 \be
 \label{eq7} a(t)= \exp\Big[\frac{-4a^2 \Delta + a^2(
 1+8b^2
 \Delta)\varepsilon}{\sigma^2}\Big]\ .
 \ee
 This expression is still exact. The attenuation factor(\ref{eq7}) is the measure of decoherence we shall use in what
follows to investigate the short--time limit of decoherence. To be
more specific, we shall place ourselves in the limit of weak
coupling between the system and the bath, $\gamma \ll \omega_0$,
and assume that the time is small compared to the relaxation time,
$\gamma t \ll 1$ and also small compared to the evolution time of
the harmonic oscillator, $\omega_0 t\ll 1$.

{\bf Uncorrelated initial conditions}. We begin by considering
initial decorrelation between the system and the heat bath.
Physically, this corresponds to the situation where  the system
and the bath are isolated prior to their coupling at $t=0$, the
system being effectively at zero temperature. In this case
$\varepsilon=0$, $\alpha(t) = N(t)$ and $\Delta(t) = C(t)$ (with
only the first double integral). In the high--temperature limit,
$T\!\gg \!\gamma,\omega_0$, we approximate the hyperbolic
cotangent in Eq.~(\ref{eqa3}) by $\mbox{\ coth }x \simeq 1/x$.
Expanding $\alpha(t)$ and $\Delta(t)$, Eqs.~(7) and (8), in lowest
order in time, we find that the attenuation coefficient
(\ref{eq7}) is given by
 \be
 \label{eq8} a_0^{\SC{FP}}(t) = \exp\Big[-\frac{4a^2T \gamma t^3}{12b^4+8 b^2  T
 \gamma t^3+ 3 t^2}\Big] \ ,
 \ee
 for the free particle ($t\!\ll \!\gamma^{-1}$), and by
 \be
 \label{eq9}
 a_0^{\SC{OH}}(t) = \exp\Big[-\frac{4a^2T
 \gamma t^3}{12 b^4+ 8b^2  T \omega_0^2 \gamma
 t^3+3  t^2}\Big] \ ,
 \ee
 for the harmonic oscillator ($t\!\ll \!\omega_0^{-1}$).
For very short times, we thus obtain the {\it same} attenuation
coefficient for both the free particle and the linear oscillator,
 \be
 \label{eq10}
 a_0(t) \simeq \exp\Big[-\frac{a^2  T \gamma
 t^3}{3b^4}\Big] \ .
 \ee
 Interference patterns between the two superposed wave packets are
hence destroyed according to a stretched exponential on a time
scale, $\tau_D\!= \!(3b^4/\gamma T a^2)^{1/3}$. The decoherence
time $\tau_D$ depends solely on the friction strength $\gamma$,
the temperature $T$ and the parameters of the initial wave packets
(separation $a$ and width $b$), and not on any system specific
quantity, like the frequency of the oscillator for example. Since
we look at times much shorter than the characteristic time of the
system, $\tau_S \sim \omega_0^{-1}$, such a system independence of
the decoherence time is to be expected. However, as we shall see,
this is only true for uncorrelated initial conditions. This result
is reminiscent of the universal regime discussed by Haake and
coworkers, where the decoherence rate was also found to be
independent of the nature of the system  \cite{haa01,haa02}. However, the
short--time attenuation factor (\ref{eq10}) does not quite belong
to this regime of very fast decoherence characterized by a
Gaussian decay law, $\exp[-(t/\tau_D)^2]$. On the other hand, the
cubic time dependence  clearly indicates that expression
(\ref{eq10}) goes beyond the (long-time) Golden Rule regime,
$t\!\gg\!\omega_0^{-1}$, and its typical exponential decay,
$\exp[-t/\tau_D]$. Here we have a much slower initial decoherence
compared to the Golden Rule expression. A similar short--time
cubic dependence of the decoherence factor already appears in
Refs.~\cite{cal83} and \cite{for01}.  We also mention that the
short-time approximation used in the derivation of
Eq.~(\ref{eq10}) amounts to neglecting the spreading of the wave
packet that appears in the denominator of the attenuation
coefficient. For instance,  the short-time spreading of the free
wave packet in Eq.~(\ref{eq8}) is given by $\la \Delta x^2(t)\ra=
b^2+ t^2/4b^2 + 2T \gamma t^3/3$, which reduces to the initial
width $b^2$ for very short times.

{\bf Thermal initial conditions}. We now turn to the case where
the system and the bath are initially correlated and in thermal
equilibrium. First, it should  be realized that thermal initial
conditions not only modify the coherence time of the system, they
also directly affect the overall coherence length \cite{haa02}. As a
matter of fact, we easily see from the coordinate representation
of the free particle equilibrium density matrix, $\la
x|\exp[-\beta p^2/2]|y\ra \simeq \exp[-(x-y)^2/2\beta]$, that
there is an exponential cutoff in $x-y$ over distances of the
order of the thermal de Broglie wavelength, $\lambda \sim
1/\sqrt{T}$. The two wave packets can therefore  only coherently
interfere if they are separated by a distance  smaller than
$\lambda$. This distance is extremely small at high temperature,
but becomes appreciable when the temperature is very low.

In the limit of high temperature, we can compute the attenuation
factor (\ref{eq7}) for the free particle, for times smaller than
the relaxation time, $t\ll \gamma^{-1}$, by expanding the
functions  $\alpha(t)$ and $\Delta(t)$ up to lowest order in $t$,
in analogy with the previous section (now keeping all the terms).
For very short times, this leads to
 \bea
 \label{eq11}
 a_{hT}^{\SC{ FP}}(t) &=& \exp\Big[ - \frac{2 a^2  T
 t^2}{4b^4+4  T b^2 t^2 +  t^2}\Big] \nonumber \\
 &\simeq& \exp\Big[-\frac{a^2 T t^2}{2 b^4}\Big]  \ .
 \eea
 The short--time expression (\ref{eq11}) is equivalent
to the result recently obtained by Ford, Lewis and O'Connell using
an exact method based on quantum distribution functions
\cite{for01b}. It is worth noticing that Ford and O'Connell have also
derived Eq.~(\ref{eq11}) with a more elementary method making only
use of basic quantum mechanics and equilibrium statistical
mechanics \cite{for01a}. This approximate method is valid in the
limit of vanishingly small friction. Equation (\ref{eq11}) shows
that, in the presence of initial thermal correlations between the
system and the bath, the short--time dependence of the exponent of
the decoherence coefficient is now quadratic. This has to be
contrasted with the cubic dependence obtained for initial
decorrelation in Eq.~(\ref{eq10}). Moreover,  the decoherence
time, $\tau_D = (2b^4/ T a^2)^{1/2}$, appears to be independent of
the friction coefficient $\gamma$. This remarkable result
indicates that at high temperature, decoherence can occur without
dissipation \cite{for01a,for01b}.

Similarly, the attenuation coefficient for the harmonic oscillator
can be calculated in the limit $t\ll \omega_0^{-1}$. Still in the
high--temperature limit, we find that for very short times it is
given by
 \bea
 \label{eq12}
 a_{hT}^{\SC{OH}}(t) &=& \exp\Big[-\frac{2a^2T
 \omega_0^2t^2 + a^2 \varepsilon(  4 b^2 T \omega_0^2 t^2)}
 {4 \omega_0^2b^4+ 4b^2  T \omega_0^2 t^2 + \omega_0^2
  t^2}\Big] \nonumber \\
 &\simeq&\exp\Big[-\frac{a^2  T t^2}{2 b^4}(1+2b^2 \varepsilon) \Big] \ ,
 \eea
 with $\varepsilon \simeq (\beta \omega_0)^2/2$.
Contrary to Eq.~(\ref{eq10}), we observe that for initial thermal
conditions, the short--time attenuation coefficients for the free
particle (\ref{eq11}) and the harmonically bound particle
(\ref{eq12}) are {\it not} identical, even for times much smaller
than the characteristic system time $\tau_S$. Equation
(\ref{eq12}) indeed contains an additional, system specific
correction, that depends on the frequency of the linear oscillator
$\omega_0$, and on the temperature $T$. This term
 finds its origin in the initial thermal
correlations existing between the system and the heat bath at
$t=0$ (see the discussion below). In the limit of high
temperature, $T\!\gg\! \omega_0$, this correction, which is
proportional to $\varepsilon$, is negligibly small. However, as we
shall see next, it becomes increasingly important as the
temperature is lowered.

In the low--temperature limit, we replace the hyperbolic cotangent
in  Eq.~(\ref{eqa3}) by unity, $\mbox{\ coth }x \simeq 1$. For
very short times, we find that the attenuation coefficient  for
the free particle reads ($t\!\ll \!\gamma^{-1}$),
 \bea
 \label{eq16}
 a_{lT}^{\SC{ FP}}(t) &=& \exp\Big[ \frac{a^2\gamma}{\pi b^4}(\ln \gamma
 t +\gamma_e-3+\ln 2) t^2\Big ]\nonumber \\
 &\simeq& \exp\Big[ \frac{a^2}{\pi b^4}\gamma t^2\ln \gamma
 t\Big ]\ ,
 \eea
where $\gamma_e$ is Euler's constant. We note that in contrast to
the high--temperature expression (\ref{eq11}), Eq.~(\ref{eq16})
now explicitly contains the damping coefficient $\gamma$ (see the recent discussion in Ref.~\cite{mar02} on this point). We also note that the exponent in Eq.~(\ref{eq16}) is actually negative since  $\gamma t< 1$. The
"$t^2\ln t$" behavior of the decoherence factor in
Eq.~(\ref{eq16}) is consistent with the result found by Romero and
Paz for the superposition of two translations \cite{rom97}. In an
analogous way, we find that the short--time expression of the
attenuation coefficient for the linear oscillator is given by ($t\!\ll \!\omega_0^{-1}$),
 \bea
 \label{eq17}
 a_{lT}^{\SC{OH}}(t) &=& \exp\Big[ \frac{2 a^2\gamma}{\pi b^4}(\ln
 \omega_0
 t +\gamma_e) t^2(1+2b^2 \varepsilon)\Big ]\nonumber \\
 &\simeq& \exp\Big[ \frac{2 a^2}{\pi b^4}\gamma t^2\ln \omega_0
 t(1+2b^2 \varepsilon)\Big ]\ ,
 \eea
 where now $\varepsilon \simeq \beta \omega_0$. In the limit of
 low temperature, $T\ll \omega_0$, $\varepsilon$ becomes very large. It
 should be noticed that the factor $\ln\gamma t$  in
 Eq.~(\ref{eq16}) comes from the first double integral in
 $C_\omega(t)$,
 Eq.~(25), whereas  the factor $\ln\omega_0 t$ in Eq.~(\ref{eq17})
 comes from the next  two single integrals of (25).
\begin{figure}[h]
\centerline{\epsfxsize=8cm
\epsfbox{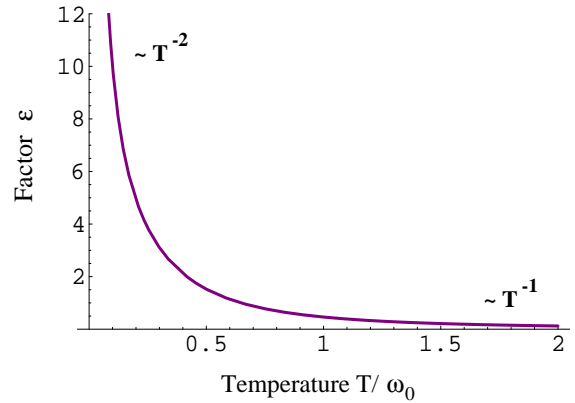}}
\vspace{1mm}
 \caption{Temperature dependence of the factor
$\varepsilon$, given by Eq.~(\ref{eq18}), in the limit of small
friction, $\gamma\ll \omega_0$.} \label{fig1}
\end{figure}
 {\bf Discussion}. From a technical point of view, the inclusion of
initial correlations between the system and the bath results in a
modification of  the integration contour in the complex--time
plane (Keldish contour) that appears in the path integral
evaluation of the influence functional \cite{mor87,mor90,gra88}. More
precisely, the effect of the initial correlations is to couple the
forward and backward integration paths along the real time axis
through an imaginary path along the Euclidean time axis, $\tau=i t
$. This leads to additional terms in the influence functional that
are given by Euclidean integrals of the form$ \int_0^\beta d\tau
f(\tau)$. The presence of the factor $\varepsilon$ in the
propagating function (\ref{eq5}) can be directly traced back to
these terms. Clearly, even for very short times, when the dynamics
of the system can be completely neglected, some of these terms,
that depend only on temperature and not on time, will still be
present. For initial thermal correlations, these terms explicitly
depend on the nature of the system through the common initial
equilibrium state of the system with the bath. As a consequence,
even for arbitrarily small times, the attenuation coefficient will
be system dependent. Only in the special case of uncorrelated
initial conditions is the attenuation factor system independent at
short times. It turns out, furthermore, that the contribution of
these system--specific terms to the decoherence time will be
negligibly small in the limit of high temperature, $\beta
\rightarrow 0$, (as easily seen from the Euclidean integral above)
and will become very important in the opposite limit of low
temperature, $\beta \rightarrow \infty$. This general discussion
confirms the results obtained for the special examples of a free
particle and a linear oscillator.

Finally, it is also interesting to  look at the temperature
dependence of the factor $\varepsilon$ which can be written in the
form
 \be
 \label{eq18}
 \varepsilon= \left[\frac{4 \gamma}{\pi\beta}\int_0^\infty
 d\omega \frac{\omega\coth(\beta
 \omega/2)}{(\omega^2-\omega_0^2)^2+4\gamma^2\omega^2}\right]^{-1}\ .
 \ee
 At high temperature, $\varepsilon$ asymptotically decays to zero
 as $T^{-1}$, whereas at low temperature it diverges as $T^{-2}$
 [see Figure (\ref{fig1})]. In the absence of damping,
 Eq.~(\ref{eq18}) reduces to the simple expression
 \be
\label{eq19}
 \varepsilon= \frac{\beta \omega_0}{\coth(\beta
 \omega_0/2)}, \hspace{1cm} \gamma=0\ .
 \ee
 We see from Eq.~(\ref{eq19}) that the high--temperature limit $\beta\rightarrow 0$
is formally equivalent to the free particle limit $\omega_0
\rightarrow 0$. This offers another explanation why the system
dependent correction to the attenuation coefficient becomes
vanishingly small at high temperature. Moreover, the presence of
the hyperbolic cotangent in the denominator of  Eq.~(\ref{eq19})
hints at a possible connection between the divergence of
$\varepsilon$ close to zero temperature and
 zero--point oscillations of the bath.

  In conclusion, we have examined the effect of initial
correlations on the short--time decoherence of a superposition of
two Gaussian wave packets. To this end, we have calculated the
attenuation coefficient for both a free particle and a linear
oscillator, with and without initial thermal correlations. We have
found that for factorizable initial conditions, the attenuation
factor, and accordingly the decoherence time, is system
independent at very short times. On the other hand, for correlated
thermal initial conditions, not only the temporal properties of
decoherence are modified ---changing  from a stretched exponential
to a Gaussian decay --- but also the coherence length is affected.
The latter is of the order of the de Broglie thermal wavelength.
Moreover, the attenuation factor now has a system dependent term,
containing the frequency of the oscillator and the temperature,
and this even at times much smaller than the system characteristic
time $\tau_S\sim \omega_0^{-1}$. The system specific correction to
the attenuation coefficient is small in the high temperature
limit, where zero--point fluctuations are neglected, but becomes
more and more important as the temperature is lowered.

 This work was funded in part by the ONR under contract N00014-01-1-0594.

\section*{Appendix A}
In this appendix we collect for convenience the exact expressions of the quantities
appearing in the propagator (\ref{eq5}). The derivation of these quantities can be
found in Ref.~\cite{mor90}.  The functions $K(t)$ and $N(t)$ are given by
 \bea
 \label{eqa1}
 K(t)&=& \nu \mbox{ cotan } \nu t \ , \\
 N(t) &=& \frac{\nu \exp[\gamma t]}{\sin \nu t} \ ,
 \eea
 where $\nu= (\omega_0^2-\gamma^2)^{1/2}$. We work in  the limit of an
underdamped oscillator where the friction coefficient $\gamma$ is
smaller than the frequency $\omega_0$ of the harmonic oscillator.
For a free particle, $\omega_0=0$, one has to replace
$\nu\rightarrow i \gamma$. The variance of the position in
equilibrium, $\kappa=\la x^2\ra_{eq}$, is given by \cite{gra88}
 \be
 \label{eqa2}
 \kappa = \frac{2 \gamma}{\pi} \int_0^\infty d\omega
 \frac{\omega \coth(\beta \omega/2)}{(\omega^2-\omega_0^2)^2+4
 \gamma^2 \omega^2}\ .
 \ee
Its asymptotic behavior at high ($ T\!\gg \! \omega_0$) and low
temperature ($T\! \ll \! \omega_0$) is respectively given by
$\kappa\sim 1/\beta \omega_0^2$ and $\kappa\sim 1/2\omega_0$. The
functions  $C(t)$ and $E(t)$ are, on the other hand, of the form
 \be
 \label{eqa3}
 f(t) = \frac{\gamma}{\pi}\int_0^{\omega_c} d\omega \,
 \omega \coth(\beta \omega/2) f_\omega(t) \ ,
 \ee
 where $f_\omega(t)$ has respectively to be replaced by
\end{multicols}
\widetext
 \bea \label{eqa4}C_\omega(t) &=& \frac{1}{\sin^2 \nu t}
\int_0^t dt'\int_0^t dt'' \sin \nu(t-t') \cos \omega (t'-t'') \sin
\nu (t-t'')
\exp[\gamma(t'+t'')] \nonumber \\
&+& \frac{4 \gamma \omega^2}{(\omega_0^2-\omega^2)^2 + 4
\gamma^2\omega^2} \frac{1}{\sin \nu t} \int_0^t dt' \cos \omega t'
\sin \nu(t-t')
 \exp[\gamma t'] \nonumber \\
&-&
\frac{2\omega(\omega^2-\omega_0^2)}{(\omega_0^2-\omega^2)^2+4\gamma^2\omega^2}
\frac{1}{\sin \nu t}\int_0^t dt'\sin \omega t' \sin\nu (t-t')
\exp[\gamma t'] +
\frac{\omega^2}{(\omega_0^2-\omega^2)^2+4\gamma^2\omega^2}\ ,\\
E_\omega(t) &=& \frac{2}{\sin \nu t}
\frac{1}{(\omega_0^2-\omega^2)^2+4\gamma^2\omega^2} \int_0^t
dt'\sin \nu(t-t') \exp[\gamma t']\Big((\omega_0^2-\omega^2) \cos
\omega t' - 2 \gamma \omega \sin \omega t' \Big)\ . \eea
\newpage
\begin{multicols}{2}
\section*{Appendix B }
In this appendix we evaluate the function $\widetilde C(t) = \sin^2 \nu t\, C(t)$ [the contributions coming from the function $E(t)$ are of higher order and can therefore be neglected in the limit considered in the paper]. 
We begin by computing the first double integral in Eq.~(25). Introducing the new variables $u=(t'+t'')/2$ and $v=t'-t''$, we have
\end{multicols}
\bea
 a_\omega(t)&=& \int_0^t dt'\int_0^t dt'' \sin \nu(t-t') \sin
\nu (t-t'')  \cos \omega (t'-t'') 
\exp[\gamma(t'+t'')]\nonumber \\
& =&\int_0^t du\int_0^t dv \sin \nu(t-u-v/2) \sin
\nu (t-u+v/2) \exp[2\gamma u] \cos \omega v \ .
\eea
The integral over $u$ is readily obtained as,
\be
\int_0^t  du\, \sin \nu(t-u-v/2) \sin
\nu (t-u+v/2) e^{2 \gamma u}= \frac{1}{4} \Big [\frac{(e^{2\gamma t}-1)
\cos\nu v}{\gamma}+\frac{\gamma \cos2 \nu t -\nu \sin 2 \nu t -\gamma e^{-2\gamma t}}{\gamma^2+ \nu^2}\Big ] \ ,
\ee
while  the integral over $v$ can be calculated using 
\be
\int_0^\infty d\omega \,\omega \coth(\beta \omega/2) \cos\omega v \simeq 2T\int_0^\infty d\omega \, \cos\omega v = 2 \pi T\, \delta(v) \ ,
\ee
at high temperature, and 
\be
\int_0^\infty d\omega\, \omega \coth(\beta \omega/2) \cos\omega v \simeq \int_0^\infty d\omega \,\omega  \cos\omega v = -\frac{1}{v^2} \ ,
\ee
at low temperature (the latter should be  interpreted as a principal value \cite{pap62}).
Furthermore, by using the following two expressions,
\bea
 &&\int_0^t \!dt' \cos \omega t' \sin \nu(t-t')e^{\gamma t'} = \frac{-\nu(\omega_0^2-\omega^2) \cos\nu t -\gamma (\omega^2+\omega^2) \sin \nu t + \nu e^{\gamma t}[(\omega_0^2-\omega^2)\cos\omega t +2 \gamma \omega \sin \omega t]}{(\omega_0^2-\omega^2)^2+4\gamma^2\omega^2}\ , \\
&&\int_0^t \!dt' \sin \omega t' \sin \nu(t-t')e^{\gamma t'}=\frac{2\gamma \nu\cos \nu t+\omega(\gamma^2-\nu^2+\omega^2)\sin \nu t+ \gamma e^{\gamma t}[-2\gamma \omega \cos\omega t +(\omega_0^2-\omega^2) \sin \omega t]}{(\omega_0^2-\omega^2)^2+4\gamma^2\omega^2}\ ,
\eea
we can rewrite the remaining three terms in Eq.~(25) in the compact form
\be
b_\omega(t) = \frac{2\omega \sin \omega t \,\nu \sin \nu t  \exp[\gamma t]-\omega^2\sin^2\nu t}{(\omega_0^2-\omega^2)^2+4\gamma^2\omega^2} \ .
\ee
The function $\widetilde C(t)$ is eventually given by
\be
\widetilde C(t) =  \frac{\gamma}{\pi}\int_0^{\omega_c} d\omega \,
 \omega \coth(\beta \omega/2) [a_\omega(t) +b_\omega(t)] \ .
\ee
\begin{multicols}{2}

\end{multicols}
\end{document}